\documentclass{desyproc}

\begin{document}
\title{Flavor changing neutral currents in top quark production and decay}


%

%

%
\author{{\slshape Efe Yazgan}  for the ATLAS, CDF, CMS, and D0 Collaborations\\[1ex]
Department of Physics and Astronomy, University of Ghent, Proeftuinstraat 86, B-9000 Ghent, Belgium }

\contribID{xy}  
\confID{7095}
\desyproc{DESY-PROC-2013-XY}
\acronym{TOP2013}
\doi            

\maketitle

\begin{abstract}
Top quark flavor changing neutral current (FCNC) interactions are highly suppressed in the Standard Model. Therefore, any large signal of FCNCs will indicate the existence of new interactions. In this paper, searches for FCNC interactions in top quark production and decay at the Tevatron and LHC are presented. FCNC searches in $t\rightarrow qZ$ and $t\rightarrow Hq$ decays, and in top quark production in $pp\rightarrow t+j$,   $pp\rightarrow t+Z$ are summarized. Effect of top quark FCNCs on single top quark cross-section, and the searches for same-sign top quark pair production through FCNCs are also described. None of the searches yielded positive results and exclusion limits on branching rations, coupling strengths and cross-sections are obtained. Future prospects of FCNC searches are also briefly discussed. 
\end{abstract}

\section{Introduction}
Flavor Changing Neutral Currents (FCNCs) are transitions that change the flavor of a fermion without changing its charge. FCNCs are forbidden at the tree level in the standard model (SM) and are suppressed at higher orders due to the GIM mechanism \cite{ref:gim}. FCNC interactions occur only at the level of quantum loop corrections with branching ratios, $\mathcal{B}(t\rightarrow Xq)\sim10^{-17}-10^{-12}$, where $X=H, \gamma, Z$ or $g$. In models beyond SM, branching ratios up to $10^{-3}$ are predicted \cite{ref:aguilar}. Therefore, any evidence of an FCNC process will indicate the existence of new physics. Searches for FCNCs might be done using specific models  (e.g. two higgs doublet model) or in a model-independent way. In this proceeding, the analysis summarized depend on model-independent methods using effective field theory approach. Assuming that the new physics involves particles with a mass scale larger than the top quark mass, the most general effective Lagrangian with terms up to dimension 5 is \cite{ref:aguilar}
\begin{align}
\label{eqn:eqn1}
-\mathcal{L}_{eff} & =  \frac{g}{2c_W}X_{qt}\overline{q}\gamma_\mu(x_{qt}^LP_L+x_{qt}^RP_R)tZ^\mu +  \frac{g}{2c_W}X_{qt}\kappa_{qt}\overline{q} (\kappa_{qt}^v+\kappa_{qt}^a\gamma_5)\frac{i\sigma_{\mu\nu}q^\nu}{m_t}tZ_\mu   \\ \nonumber
 & +  e\lambda_{qt}\overline{q}(\lambda_{qt}^v+\lambda_{qt}^a\gamma_5)\frac{i\sigma_{\mu\nu}q^\nu}{m_t}tA^\mu+g_s\zeta_{qt}\overline{q}(\zeta_{tq}^v+\zeta_{qt}^a\gamma_5)\frac{i\sigma_{\mu\nu}q^\nu}{m_t}T^aqG^{a\mu}  & \\ \nonumber
 & +  \frac{g}{2\sqrt{2}}g_{qt}\overline{q}(g_{qt}^v+g_{qt}^a\gamma_5)tH + h.c. &  \nonumber
\end{align}
\noindent where $q^\nu=(p_t-p_q)^\nu$ with $p_t$ and $p_q$ representing four-momentum of the top quark and b quark, respectively. The symbols $\overline{q}$ and $t$ represent the quark fields. 
The coupling constants are normalized as $|x_{qt^L}|^2+|x_{qt}^R|^2=1$, $|\kappa_{qt}^\nu|^2+|\kappa_{qt}^a|^2=1$, etc., with $X_{qt}$, $\kappa_{qt}$, $\lambda_{qt}$, $\zeta_{qt}$, and $g_{qt}$ $\in\mathbb{R^+}$
and h.c. represents Hermitian conjugate. For more details, see  \cite{ref:aguilar}.
Implementation of each term may be different for some of the measurements presented here, and therefore exclusion limits are not directly comparable without the necessary replacements for different representations. The limits on the couplings in this paper, are given with the notation in their corresponding publications.


\section{FCNCs in top quark decays in $t\overline{t}$ events}
\subsection{$t\rightarrow Zq$ decays}
\begin{figure}[h]
\begin{center}
\begin{tabular}{ccc}
\includegraphics[width=0.3\textwidth]{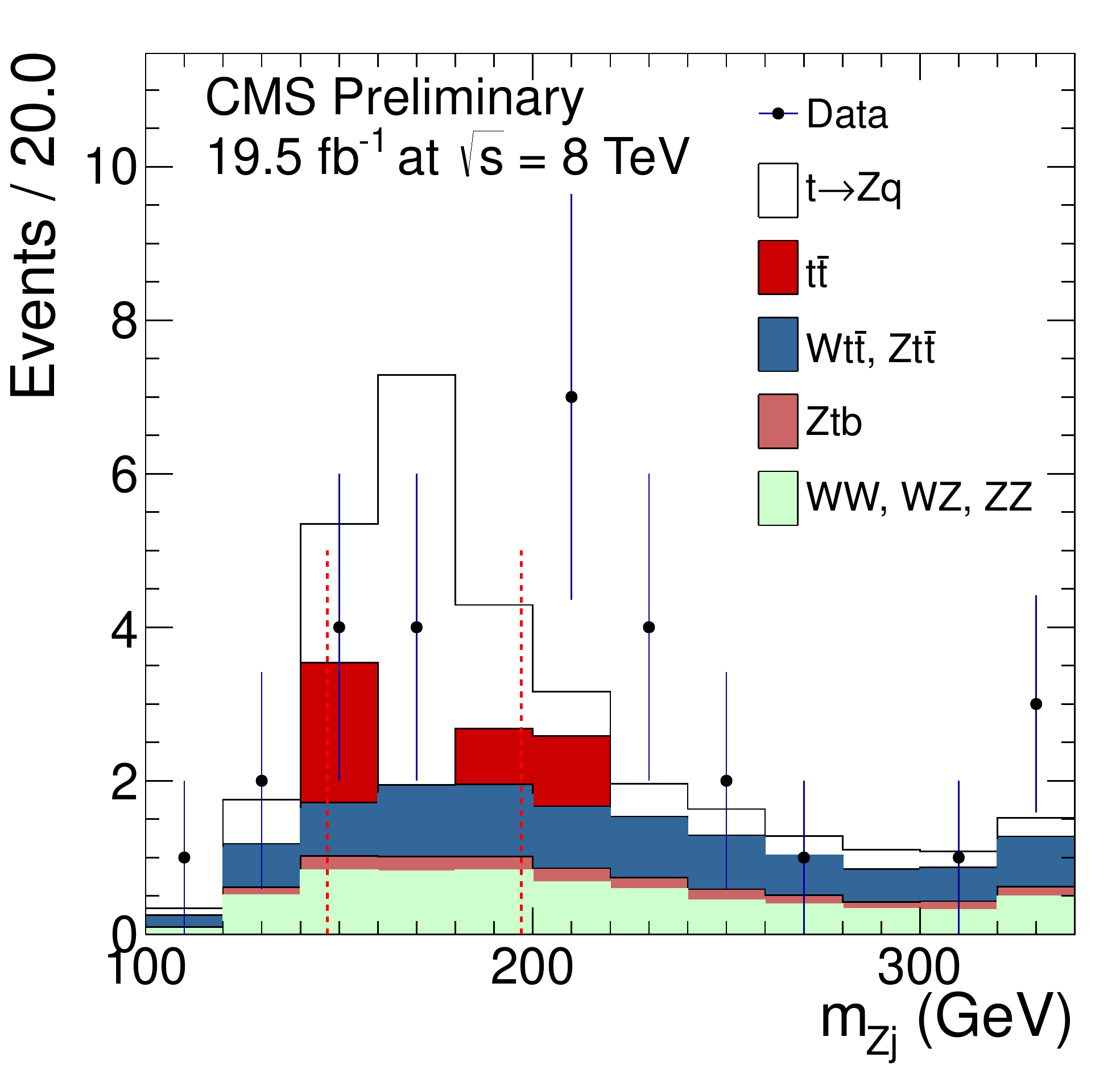} & 
\includegraphics[width=0.3\textwidth]{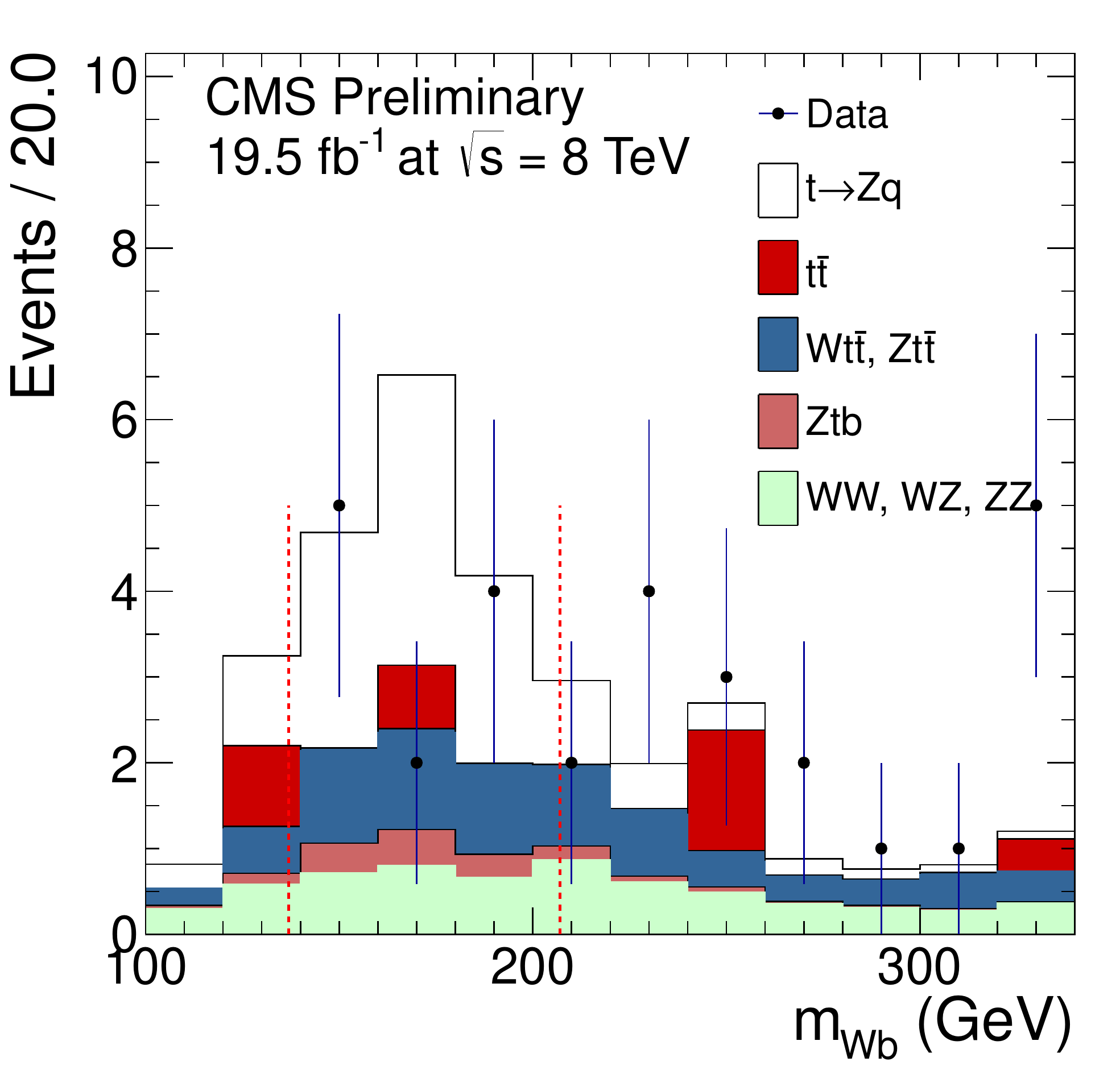} & 
\includegraphics[width=0.3\textwidth]{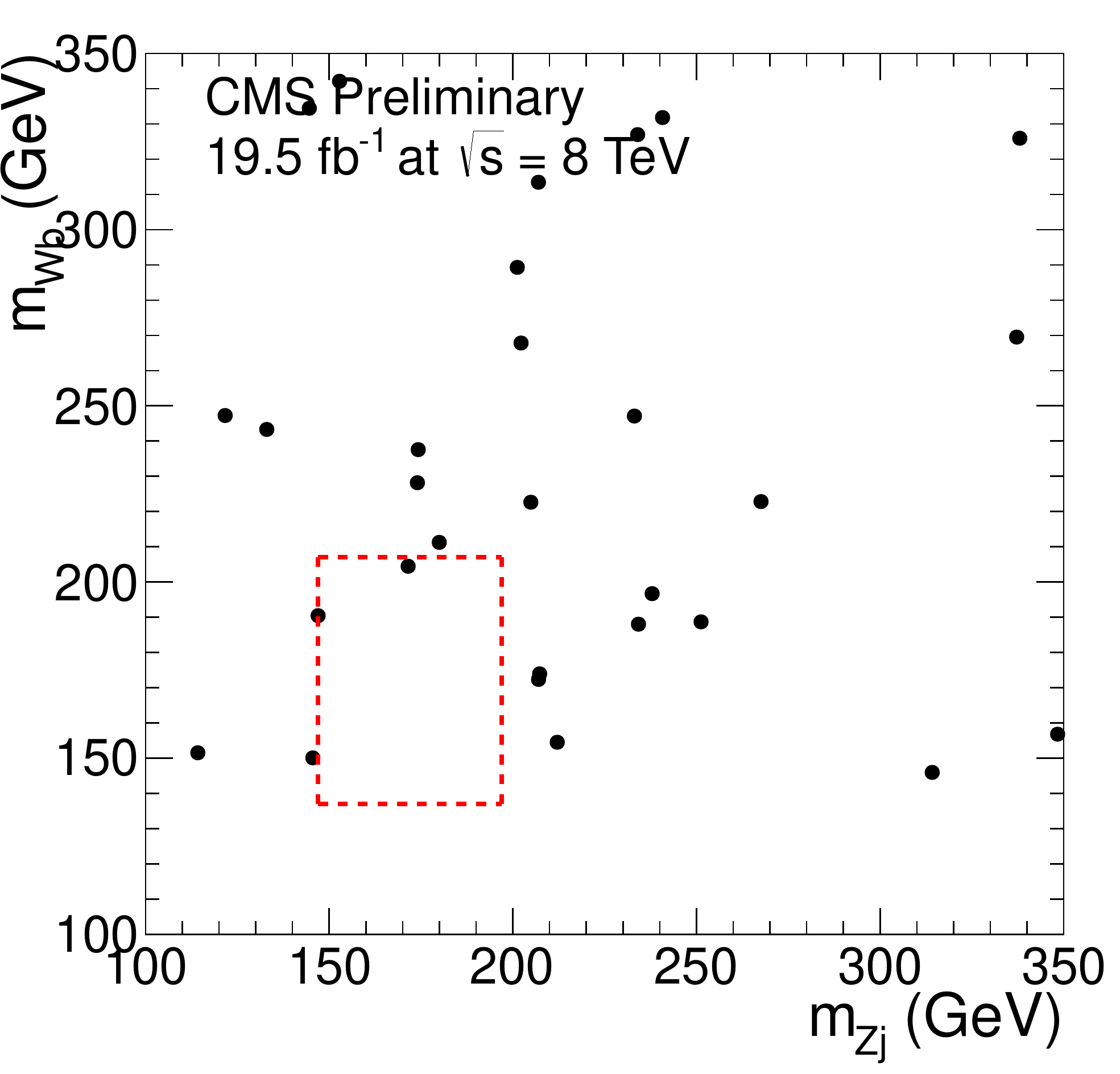} 
\end{tabular}
\end{center}
\caption{Comparison between data and MC distributions of the $m_{Zj}$ (left plot), $m_{Wb}$ (middle plot) and the data points on the $m_{Zj}$ vs $m_{Wb}$ plane (right plot). For the 2D scatter plot the data points are shown before the top quark mass selection requirements. Top quark mass requirements are shown as  dotted vertical lines in the left and the middle plots and as a dotted box on the right plot. The expected signal distributions are normalized so that $\mathcal{B}(t\rightarrow Zq)=0.1\%$.  
}
\label{Fig1}
\end{figure}

The exclusion limits obtained from searches for FCNCs processes in top quark decays in $t\overline{t}$ events are summarized in Table~\ref{tab:tqz}. 
Note that there is no  published $t\rightarrow \gamma q$ result at the LHC yet.  
The most precise exclusion limit on the $t\rightarrow Zq$ branching ratio is obtained by CMS using 19.7 fb$^{-1}$ proton-proton collision data at a center-of-mass energy ($\sqrt{s}$) of 8 TeV \cite{Chatrchyan:2012a}.  
The CMS analysis is made in the $t\overline{t}\rightarrow Wb+Zq\rightarrow\ell\nu b+\ell\ell q$ final state. Three lepton events are selected with the additional requirements of large missing transverse energy, at least two jets among which exactly one is required to be b-tagged. For signal a MadGraph \cite{Alwall:2011}+Pythia \cite{Sjostrand:2006} sample is used and backgrounds are estimated using a data-driven approach. The selected $Z$ boson and the jet as well as W boson and b-tagged jet are paired to reconstruct the top quarks. 
After all selections, the signal region is defined by a 35 GeV $Wb$  and 25 GeV $Zj$ mass window around the top quark mass. The signal, background and data distributions are shown in Fig.~\ref{Fig1}. The signal region contains one event while the expected SM background is 3.1$\pm$5.1 events. Therefore, there is no excess of events over the SM background. 
The process under investigation can be represented by the first term of the effective Lagrangian in Eq.~\ref{eqn:eqn1}. 
A branching ratio, $\mathcal{B}(t\rightarrow Zq)>0.07\%$ is excluded at the 95\% confidence level (CL). The expected 95\% CL upper limit is 0.11\%. Combined with the search at  7 TeV \cite{Chatrchyan:2012e}, the limit is $\mathcal{B}(t\rightarrow Zq)>0.05\%$. 
The dominant systematic uncertainties in this measurement are factorization and renormalization scales, parton distribution functions (PDFs) and the $t\overline{t}$ cross-section. 

\begin{table}
\centerline{\begin{tabular}{|l|l|l|l|l|l|r|}
\hline
  $\sqrt{s}$ & Detector  & Channel & $\mathcal{B}(t\rightarrow Zq)$ & $\mathcal{B}(t\rightarrow \gamma q)$ & Reference\\
 (TeV) & (integrated luminosity) & & (\%) & (\%) & \\
\hline
 1.8                & CDF ($\sim$110 pb$^{-1}$) & dilepton & 33 & 3.2 & \cite{Abe:1997fz} \\
 1.96                & CDF (1.9 fb$^{-1}$) & dilepton & 3.7 & - & \cite{Aaltonen:2008ac} \\
 1.96                & D0 (4.1 fb$^{-1}$) & trilepton & 3.2 & - & \cite{Abazov:2011qf} \\
 7                & ATLAS (2.1 fb$^{-1}$) & trilepton & 0.73 & - & \cite{Aad:2012ij} \\
 7                & CMS (5 fb$^{-1}$) & trilepton & 0.21 & - & \cite{Chatrchyan:2013a} \\
 7 + 8                & CMS (5.0 + 19.7 fb$^{-1}$) & trilepton & 0.05 & - & \cite{Chatrchyan:2012a} \\
\hline
\end{tabular}}
\caption{Observed branching ratio exclusion limits for $t\rightarrow Zq$ and $t\rightarrow \gamma q$ decays in $t\overline{t}$ events at 95\% C.L.}
\label{tab:tqz}
\end{table}

\newpage

\subsection{$t\rightarrow Hq$ decays}
The discovery of the Higgs boson by the ATLAS \cite{Aad:2012higgs} and CMS \cite{Chatrchyan:2012higgs} collaborations allows us to search for FCNC interactions occurring through the mediation of the Higgs boson. 
The ATLAS collaboration conducted a search for FCNC in the $t\rightarrow cH$ decays with $H\rightarrow\gamma\gamma$ using 4.7 fb$^{-1}$ and 20.3 fb$^{-1}$ data collected at $\sqrt{s}=$ 7 and 8 TeV, respectively \cite{Aad:2013a}.  
The search is made using $t\overline{t}$ events for which one of the top quarks decays to $cH$ and the other to $bW$. Both hadronic and leptonic decays of the $W$ bosons are considered. Backgrounds for non-resonant $\gamma\gamma$ final state are found to be small after  $t\overline{t}$ event selection. \begin{wrapfigure}{r}{0.5\textwidth}
\centerline{\includegraphics[width=0.5\textwidth]{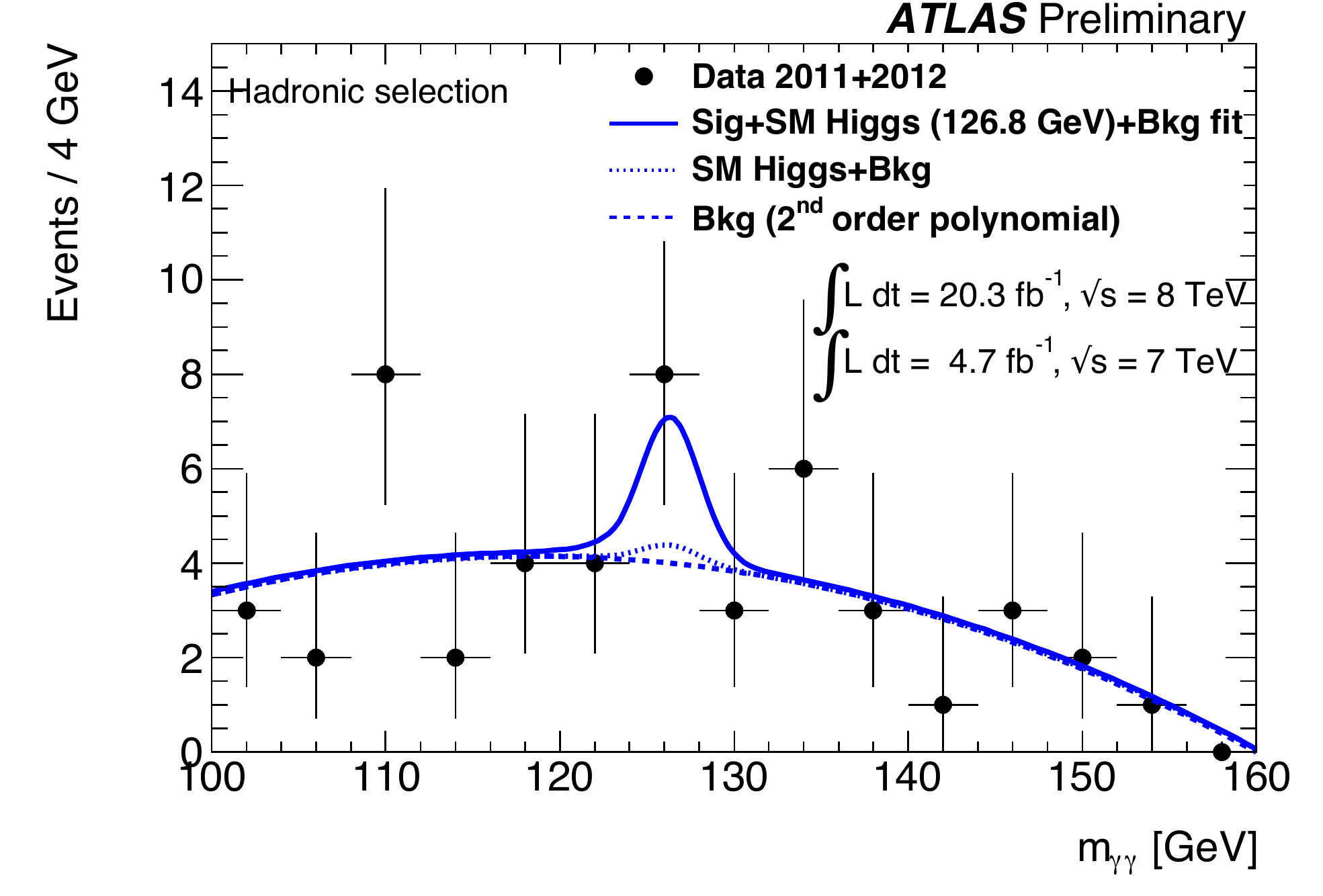}}
\caption{The diphoton mass spectrum using the selected events in the hadronic channel.}\label{Fig2}
\end{wrapfigure}
The signal signature is two high $E_T$ well identified and isolated photons. The hadronic channel is selected by requiring at least four jets with at least 1 b-tagged jet and the non-existence of leptons in the event. The leptonic channel is defined by exactly one lepton and high transverse mass defined by the lepton and $E_T^{miss}$. In addition, the events are required to fall in certain $m_{\gamma\gamma j}$ and $m_{jjj}$ mass windows. The backgrounds include SM Higgs boson backgrounds from gluon fusion, vector boson fusion (qqH), Higgs-strahlung associated production (WH, and ZH), associated Higgs boson production with a $t\overline{t}$ pair ($t\overline{t}H$), and tH production. Moreover, non-resonant two-photon production with up to three partons is also considered in the backgrounds. 
A maximum likelihood fit performed on the selected data (50 events in the hadronic channel and 1 event in the leptonic channel) yielded a total of 3.7$^{+4.4}_{-3.7}$ signal events. 
The diphoton mass spectrum using the selected events in the hadronic channel is shown in Figure \ref{Fig2}. 
Assuming $M_H=126.8$ GeV, the observed limit on the branching ratio is 0.83\% at the 95\% C.L. and the upper bound on the $\lambda_{tcH}$ coupling of 0.17. 

Craig et al. \cite{Craig:2012} obtained a branching ratio limit of 2.7\% at 95\% C.L. for $M_H=125$ GeV re-interpreting a CMS anomalous multi-lepton search conducted at $\sqrt{s}=$ 7 TeV \cite{Chatrchyan:2012b}.  At $\sqrt{s}=$ 8 TeV, CMS searched for $t\rightarrow cH$ decays from the $H\rightarrow WW^*\rightarrow\ell\nu\ell\nu$, $H\rightarrow\tau\tau$, $H\rightarrow ZZ^*\rightarrow jj\ell\ell,\nu\nu\ell\ell,\ell\ell\ell\ell$ processes in $t\overline{t}$ production \cite{Chatrchyan:2013b}. The searches have been made in exclusive multi-lepton channels defined by lepton charge flavor combinations, $E_T^{miss}$, jet activity, consistency of the invariant mass of opposite-sign lepton pairs with a Z boson, and the presence of b-jets and taus. 
It is found that the signal is more dominant in the three lepton channels with no hadronic $\tau$ particles, no opposite-sign same-flavor pair or an opposite-sign same-flavor pair off Z boson, and a b-tagged jet. No excess is observed over the SM backgrounds. 
The limits on the branching ratios are displayed in Table \ref{tab:tcH} for each decay channel. For the CMS branching ratio limits the assumed Higgs boson mass is  $M_H=125.5$ GeV. The combined branching fraction limit at 95\% C.L. obtained by CMS is 1.28\% (also shown in Table \ref{tab:tcH}). Note that the exclusion limits were revised after $TOP2013$ conference. The limit obtained by CMS is complementary to the ATLAS search in the $H\rightarrow\gamma\gamma$ decay channel.

\begin{table}
\centerline{\begin{tabular}{|l|l|l|l|l|}
\hline
  $\sqrt{s}$ & Detector & Decay mode  & $\mathcal{B}(t\rightarrow cH)$ & Ref. \\
     (TeV)            & (integrated luminosity) & & (\%) &  \\
\hline
7+8  & ATLAS (4.7+20.3 fb$^{-1}$) & $H\rightarrow\gamma\gamma$ &  0.83 & \cite{Aad:2013a} \\
8  & CMS (19.5 fb$^{-1}$) & $H\rightarrow WW$  &  1.58 & \cite{Chatrchyan:2013b} \\
8  & CMS (19.5 fb$^{-1}$) & $H\rightarrow\tau\tau$ &  7.01 & \cite{Chatrchyan:2013b} \\
8  & CMS (19.5 fb$^{-1}$) & $H\rightarrow ZZ$ &  5.31 & \cite{Chatrchyan:2013b} \\
8  & CMS (19.5 fb$^{-1}$) & $H\rightarrow WW+\tau\tau+ZZ$ &  1.28 & \cite{Chatrchyan:2013b} \\
\hline
\end{tabular}}
\caption{Observed branching ratio limits at 95\% C.L. for FCNC process in $t\rightarrow cH$ decays in $t\overline{t}$ events.}
\label{tab:tcH}
\end{table}

\section{FCNCs in top quark production in single top events}

It is difficult to distinguish the $t\rightarrow gq$ final state from the QCD multi-jets background. Instead, a much higher sensitivity can be achieved in the searches for the anomalous single top production via the $qg\rightarrow t$ process. In the final state, a quark, a gluon, or a Z boson can accompany the top quark. In the following, we summarize the searches made for anomalous top quark production in $pp\rightarrow t$, 
$pp\rightarrow t+q/g$, and $pp\rightarrow t+Z$ processes. The branching ratio exclusion limits obtained from these searches are summarized in Table~\ref{tab:singletop}. 
\begin{table}[h]
\centerline{\begin{tabular}{|l|l|l|l|l|l|l|}
\hline
  $\sqrt{s}$ & Detector  &  $\mathcal{B}(t\rightarrow gu)$ & $\mathcal{B}(t\rightarrow gc)$ & $\mathcal{B}(t\rightarrow Zu)$ & $\mathcal{B}(t\rightarrow Zc)$ & Ref.\\
 (TeV) & (integrated luminosity) & (\%) & (\%) &(\%)  & (\%) &  \\
\hline
 1.96                & CDF (2.2 fb$^{-1}$) & 0.039 & 0.57 & - & - &\cite{Aaltonen:2009a} \\
 1.96                & D0 (2.3 fb$^{-1}$) & 0.02 & 0.39 & - & - & \cite{Abazov:2010a} \\
 7                & ATLAS (2.1 fb$^{-1}$) & 0.0057 & 0.027 & - & - & \cite{Aad:2012a} \\
 7                & CMS (4.9 fb$^{-1}$) & 0.56 & 7.12 & 0.51 & 11.40 & \cite{Chatrchyan:2012c} \\
 8                & ATLAS (14.2 fb$^{-1}$) & 0.0031 & 0.016 & - & - & \cite{Aad:2013b} \\
\hline
\end{tabular}}
\caption{Observed branching ratio limits for FCNC process in single top production.}
\label{tab:singletop}
\end{table}
\subsection{$pp\rightarrow t$}
The main differences of $qg\rightarrow t$ from the SM processes are that the top quark is produced with  almost zero $p_{T}$ and therefore $W$ and the b-jet are almost back-to-back; the $p_{T}$ of the W boson is larger than that of V+jets and diboson and therefore the decay products of the W boson have small opening angles. Another difference is the different charge asymmetry in the two cases.  \begin{wrapfigure}{r}{0.4\textwidth}
\centerline{\includegraphics[width=0.4\textwidth]{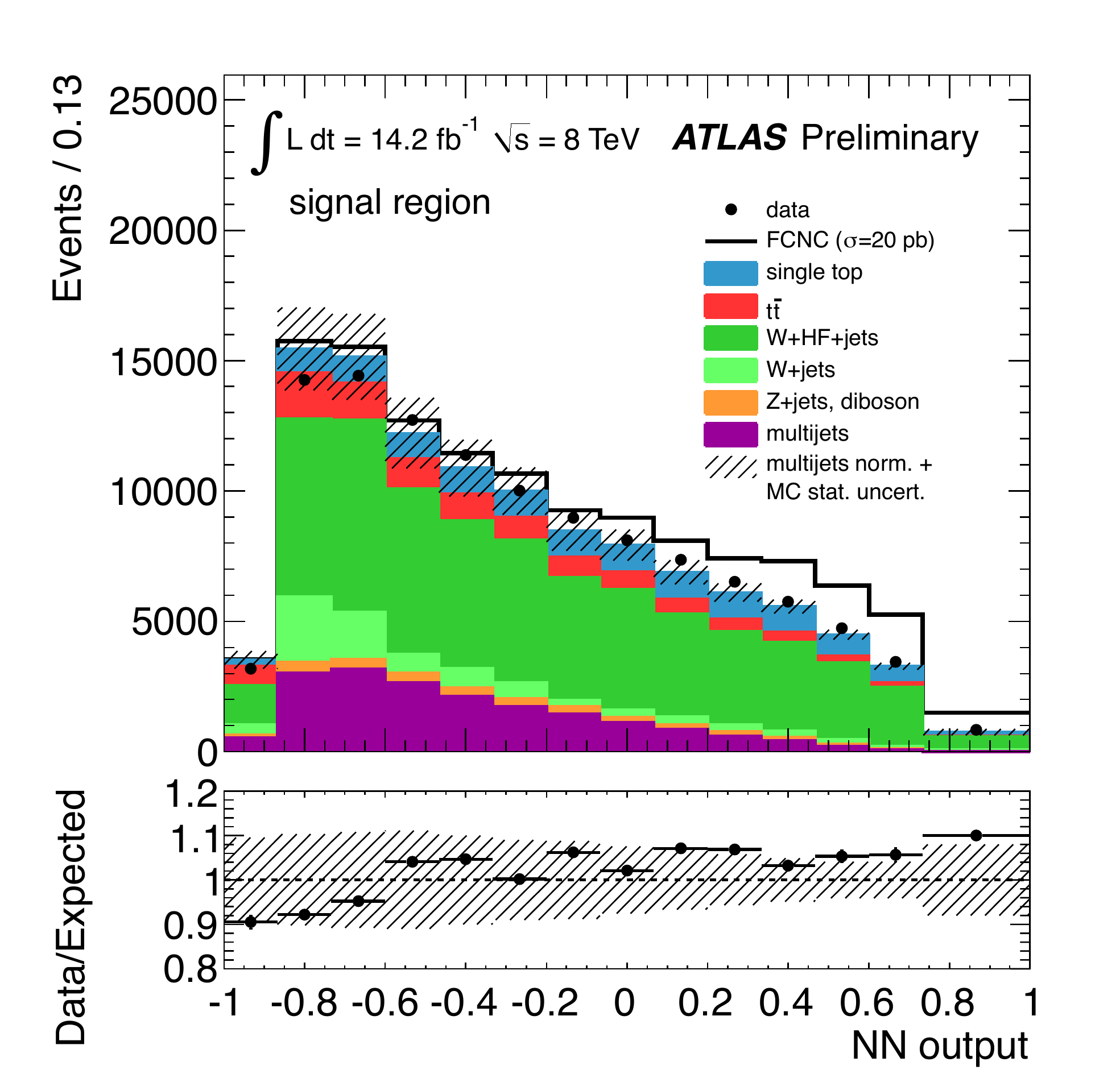}}
\caption{Output distribution of the neural network in the signal region. The signal distribution, scaled to a cross section of 20 pb, is stacked on top of the backgrounds.}\label{fig:yazgan_efe_fig3}  
\end{wrapfigure} ATLAS and CDF collaborations made searches for this process \cite{Aaltonen:2009a, Aad:2012a, Aad:2013b}. The signal simulation is made using $PROTOS$ \cite{ref:aguilar2010} and $TOPREX$ \cite{Slabospitsky:2002} for the ATLAS ($\sqrt{s}=$7 TeV) and CDF analysis respectively.  For the measurement at $\sqrt{s}=$8 TeV, ATLAS used a new generator, $ME_{TOP}$ \cite{Coimbra:2012}. This event generator provides the calculation for the FCNC process at approximate next-to-leading order level. Both collaborations used Bayesian Neural Networks (BNN) to discriminate signal and backgrounds which are dominated by $W+jets$ and $QCD~multijets$. \begin{wrapfigure}{r}{0.4\textwidth}
\centerline{\includegraphics[width=0.37\textwidth]{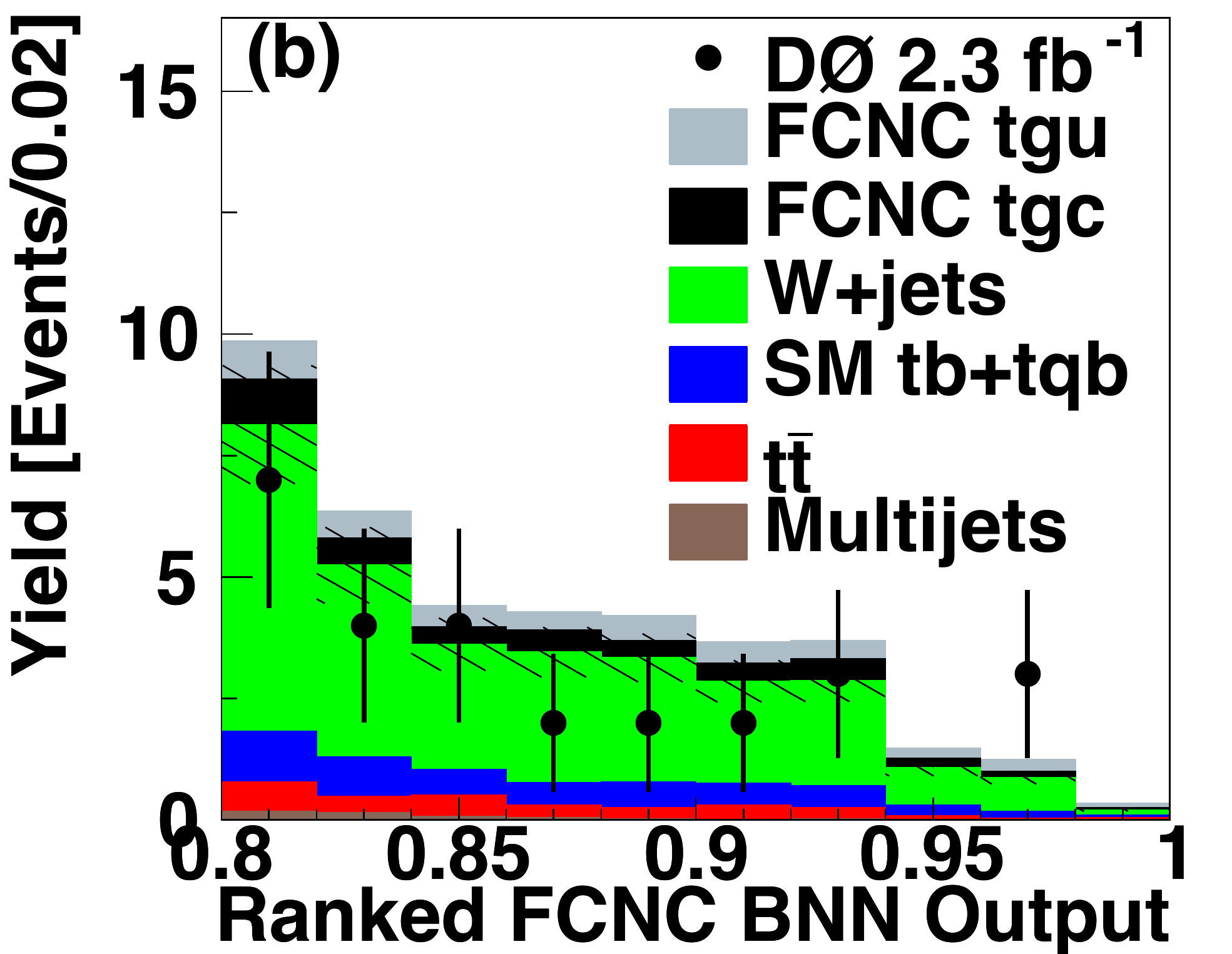}}
\caption{Background distributions and data for the FCNC BNN output at the high discriminant region.}\label{fig:azgan_efe_fig4}  
\end{wrapfigure}Binned maximum likelihood fit to the BNN output distributions are performed. BNN output distributions normalized to the binned maximum likelihood fit results and with the signal scaled to 20 pb obtained from the ATLAS analysis are shown in Figure~\ref{fig:yazgan_efe_fig3}. The process that is being searched is represented by the the fourth term of  Eq.~\ref{eqn:eqn1}. The best branching ratio exclusion limits are $B(t\rightarrow u+g)<3.1\times10^{-5}$ and $B(t\rightarrow c+g)<1.6\times10^{-4}$ obtained by ATLAS using 14.2 fb$^{-1}$ of $\sqrt{s}=$8 TeV data \cite{Aad:2013b}. From this analysis, the 95\% C.L. upper limit on the production cross section is determined to be 2.5 pb and the upper limits on the coupling constants are $\kappa_{ugt}/\Lambda<5.1\times10^{-3}$ TeV$^{-1}$ (assuming $\kappa_{cgt}/\Lambda=0$) and $\kappa_{cgt}/\Lambda<1.1\times10^{-2}$ TeV$^{-1}$ (assuming $\kappa_{ugt}/\Lambda=0$). For these exclusion limits, the dominant systematic uncertainties are the jet energy scale and resolution, b-jet tagging efficiency and parton distribution functions (PDFs).  

\subsection{$pp\rightarrow t+q/g$}
The D0 experiment made a search for the production of a top quark with an additional jet using 2.3 fb$^{-1}$ of data \cite{Abazov:2010a}. The final state is similar to the t-channel SM single top quark production. The dominant background for this process is $W+jets$. The signal and background separation is obtained by BNNs. The signal and single top quark backgrounds are simulated by $SINGLETOP$ MC \cite{Boos:2004,Boos:2006}.  For BNN, 54 variables adopted from a subset of the single-top measurement variables and variables from a previous FCNC analysis \cite{Abazov:2007a}. Discriminating variables are individual object and event kinematics, top quark reconstruction, jet width, and angular correlations. Fig.~\ref{fig:azgan_efe_fig4} displays the background distributions (normalized to their observed limits) and data for the combined BNN discriminants. No FCNC signal is observed and branching ratio limits of $2.0\times10^{-4}$ for $tgu$ and $3.9\times10^{-3}$ for $tgc$ vertices are obtained. The process is represented by the fourth term of Eq.~\ref{eqn:eqn1}. Upper limits on couplings are $\kappa_{tgu}/\Lambda<0.013$ TeV$^{-1}$ and $\kappa_{tgc}/\Lambda<0.057$ TeV$^{-1}$ and the upper limits on the cross-sections are 0.20 and 0.27 pb for the $tgu$ and $tgc$ vertices. 
Dominant systematic uncertainties are jet energy scale and b-jet tagging modeling. 

\subsection{$pp\rightarrow t+Z$}
\begin{figure}[h]
\begin{tabular}{cc}
\includegraphics[width=0.45\textwidth]{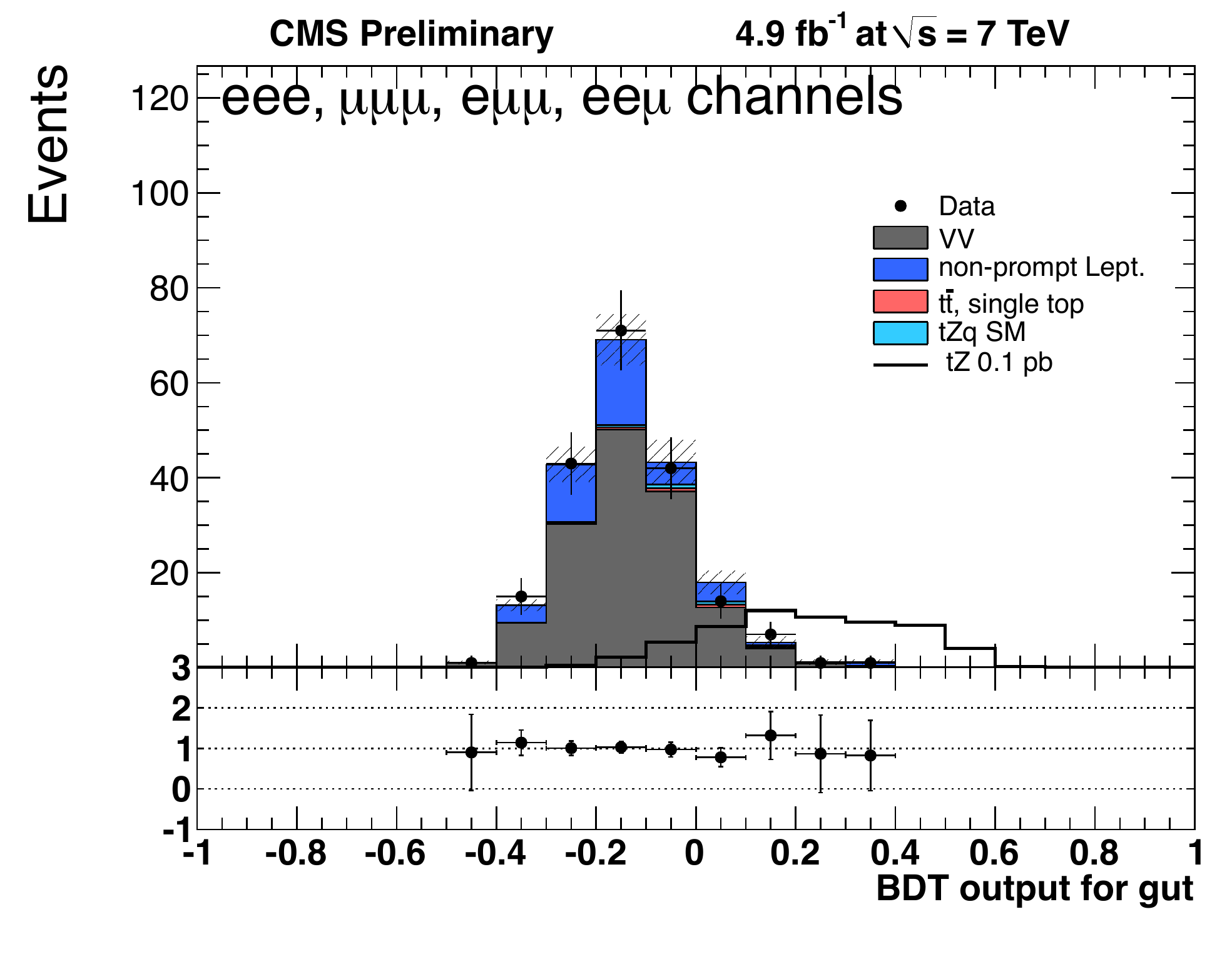} & 
\includegraphics[width=0.45\textwidth]{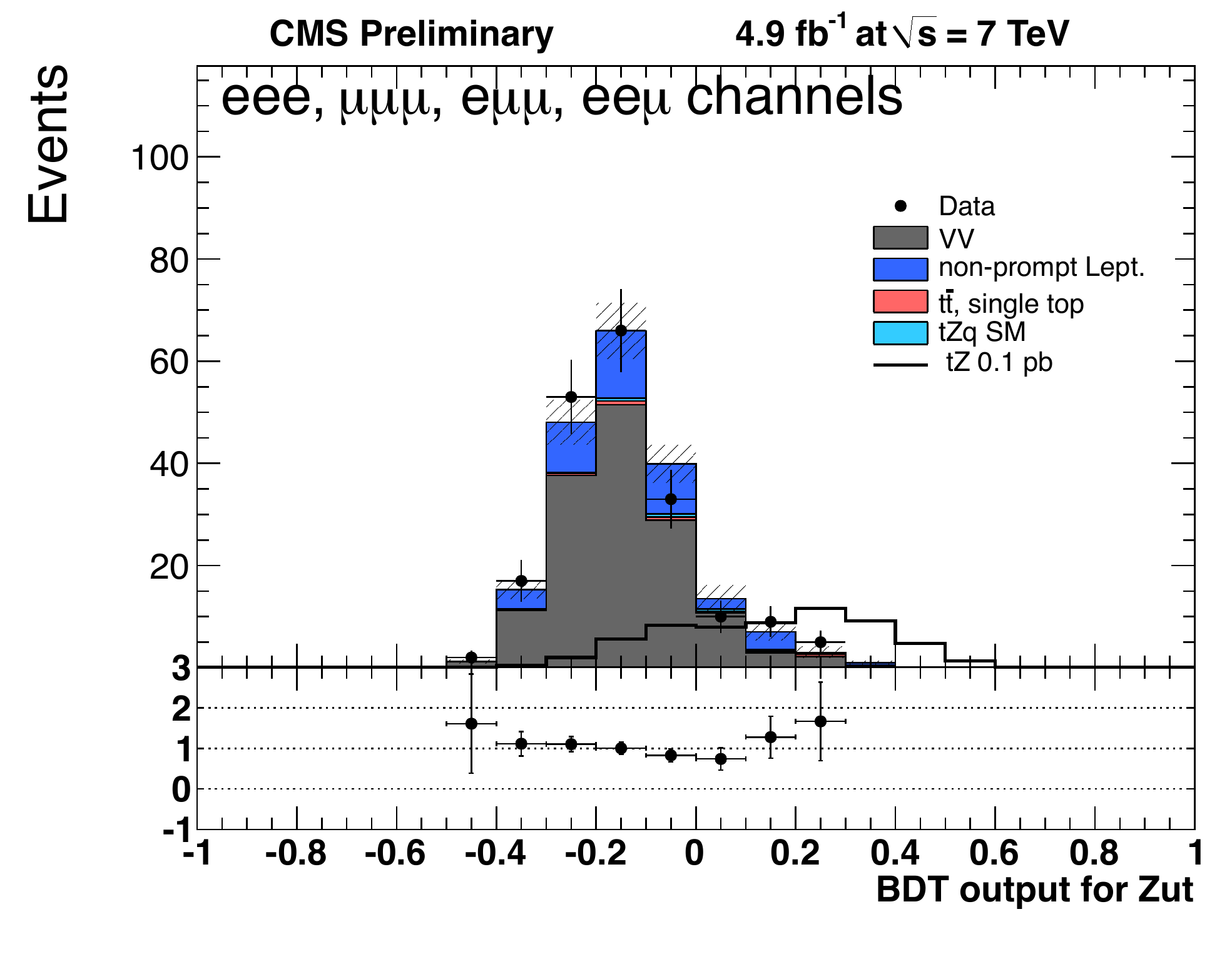} 
\end{tabular}
\caption{BDT output distribution for the $gut$ coupling (left plot) and $Zut$ (right plot), summed for the four tri-lepton channels. Total uncertainty is shown as hatched areas. The FCNC signal is normalized to a cross-section of 0.1 pb. }\label{FigSingleJeremy}
\end{figure}

CMS performed a search for the FCNC process in single top quark production in association with a Z boson using a 5 fb$^{-1}$ data sample at $\sqrt{s}=7$ TeV. In the analysis both $gqt$ and $Zqt$ vertices are probed simultaneously unlike the standard single top quark FCNC searches. The model described in \cite{Agram:2013} is used and the probed vertices are described by the second and fourth terms of Eq.~\ref{eqn:eqn1}.  The $Zqt$ vertex is also probed by the searches in top quark decays as described in the previous sections.   
The signal signature is three isolated leptons and a b-tagged jet. The signal simulation is made using MadGraph+Pythia. The signal is extracted using kinematic variables and b-jet tagging information combined using a Boosted Decision Tree (BDT). The main backgrounds are fake leptons  from the $Z+jets$ process. Other backgrounds are $ZZ+jets$, $t\overline{t}$, and $tZq$.The BDT shapes are taken from data for $Z+jets$ inverting the third lepton isolation and low $E_T^{miss}$ and other shapes are taken from simulation. 
Figure \ref{FigSingleJeremy} displays the BDT output distribution for the $gut$ coupling, summed for the four tri-lepton channels.
No FCNC signal is observed and upper limits are derived. The limits on the branching ratios are listed in Table \ref{tab:singletop}. The observed upper limits on the coupling strengths are  $\kappa_{gut}/\Lambda<0.10$ TeV$^{-1}$, $\kappa_{gct}/\Lambda<0.35$ TeV$^{-1}$, $\kappa_{zut}/\Lambda<0.45$ TeV$^{-1}$, and $\kappa_{Zct}/\Lambda<2.27$ TeV$^{-1}$. 

\newpage
\clearpage

\subsection{Single Top Quark t-channel cross section}
\begin{wrapfigure}{r}{0.5\textwidth}\centerline{\includegraphics[width=0.4\textwidth]{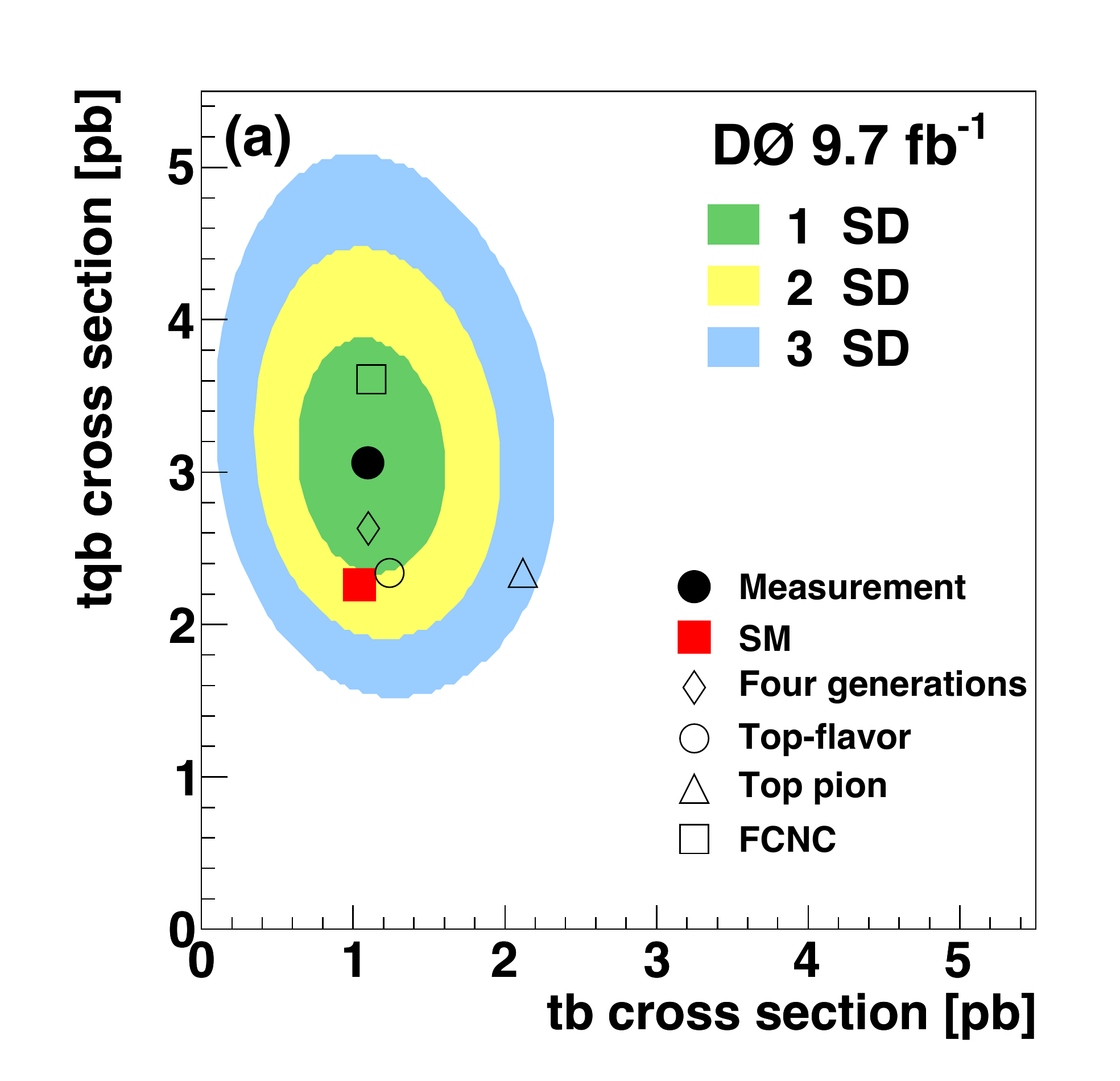}}
\caption{Single top production cross-section in the s- vs. t-channel plane. The sensitivity to different models is also shown.}\label{FigSingleTopCS}
\end{wrapfigure}FCNC modifies the t-channel production rate \cite{Tait:2001}. Figure \ref{FigSingleTopCS} shows the s- and t-channel cross-section measurement displaying the equal probability contours for the discriminant with one, two, and three standard deviations \cite{D0_singletop}. The figure also shows the prediction from SM and from different specific new physics models that can modify the s- or t-channel cross-section. One of the models shown is an FCNC model that assumes a coupling of $k_{tug}/\lambda=0.036$ \cite{Tait:2001, Abazov:2007a} modifying the SM t-channel cross-section. The D0 measurement is consistent with the SM however to exclude the FCNC model with the assumed parameters, more data is needed. The dominant systematic uncertainties are multijet normalization, $W/Z+jets$ heavy flavor correction,  ISR/FSR, $t\overline{t}$ cross-section, and b-jet tagging.\\

\section{Same-sign top quark pair production}
\begin{wrapfigure}{r}{0.5\textwidth}
\begin{tabular}{c}
\centerline{\includegraphics[width=0.5\textwidth]{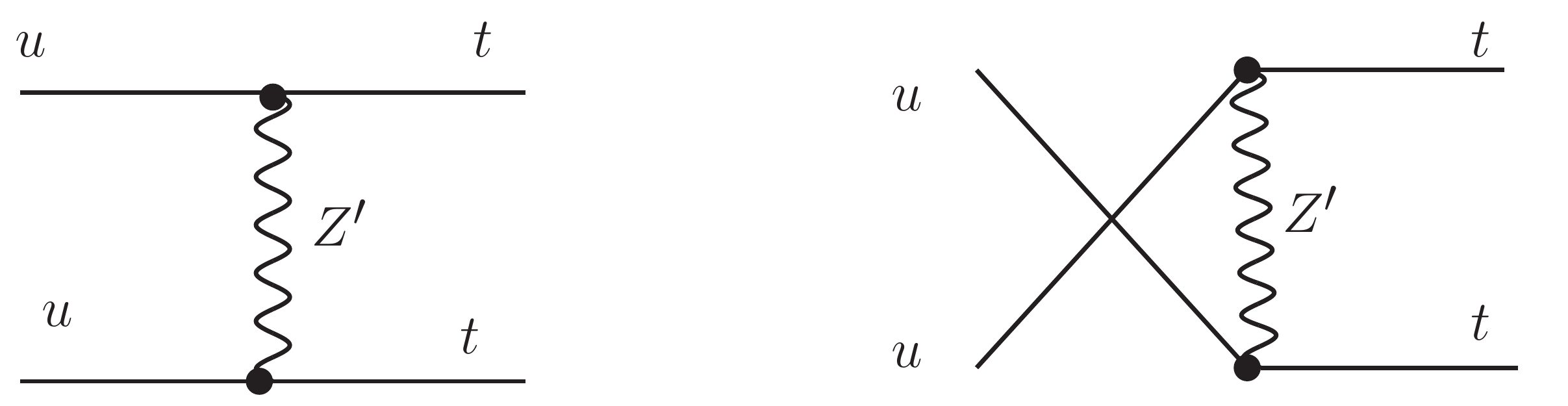}} \\
\centerline{\includegraphics[width=0.5\textwidth]{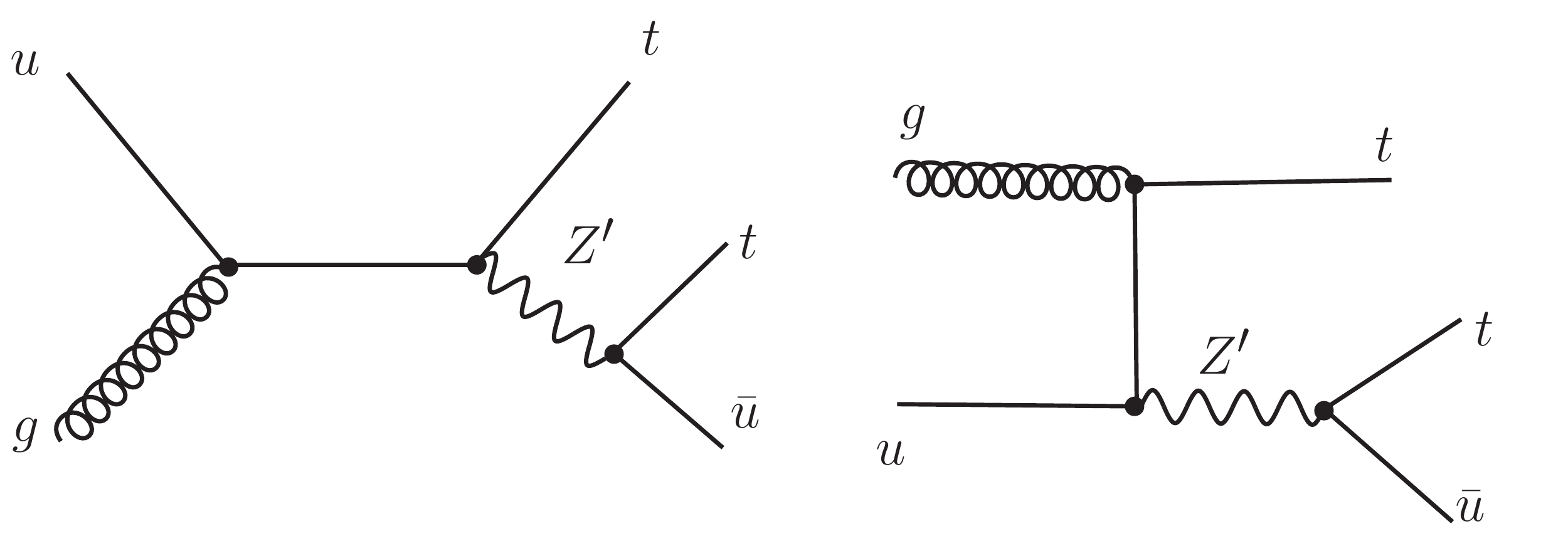}}
\end{tabular}
\caption{Diagrams for $t\overline{t}$ (top) $t\overline{t}j$ (bottom) production mediated by a $Z'$.}\label{FigSameSign}
\end{wrapfigure}The existence of same-sign top quark pair production may indicate the existence of a new heavy resonance. 
The search for same-sign top quark pair production is motivated by the  fact that the models to explain the $t\overline{t}$ forward-backward asymmetry ($A_{FB}$) observed at the Tevatron \cite{Abazov:2008a, Aaltonen:2008a,Aaltonen:2011a, Abazov:2011a, Aaltonen:2013a,Aaltonen:2013b,Aaltonen:2013c,Abazov:2013a, Abazov:2013b} usually involve FCNCs mediated by a new massive $Z'$ boson. 
The searches for top quark pair production are made looking for same-sign dilepton events by CDF \cite{Aaltonen:2011b}, CMS \cite {Chatrchyan:2012d}, and ATLAS \cite{Aad:2012b, Aad:2013c} experiments.  
 As shown in Figure \ref{FigSameSign}  by CMS and ATLAS collaborations, the non-existence of same-sign top quark production indicates that the FCNC interpretation of the Tevatron $A_{FB}$ is disfavored. The most stringent limit is obtained by the ATLAS collaboration using 14.3 fb$^{-1}$ of $\sqrt{s}=8$ TeV $pp$ collision data \cite{Aad:2013c}.
The signal signature is same-sign dilepton events accompanied by jets in which at least one of them is  a b-jet. Moreover, a missing transverse energy of 40 GeV, and an $H_T$ of 550 GeV are required. The signal simulation is made using the PROTOS event generator. The dominant backgrounds are misidentified leptons, charge  misidentifications, and $ttW+jets$. None of the searches by the different experiments yielded positive results. The 95\% C.L. exclusion limits on the cross section and couplings are obtained by the ATLAS collaboration are shown in Table~\ref{tab:samesigntop} for different chirality configurations. 
\begin{figure}
\begin{center}
\begin{tabular}{cc}
\includegraphics[width=0.57\textwidth]{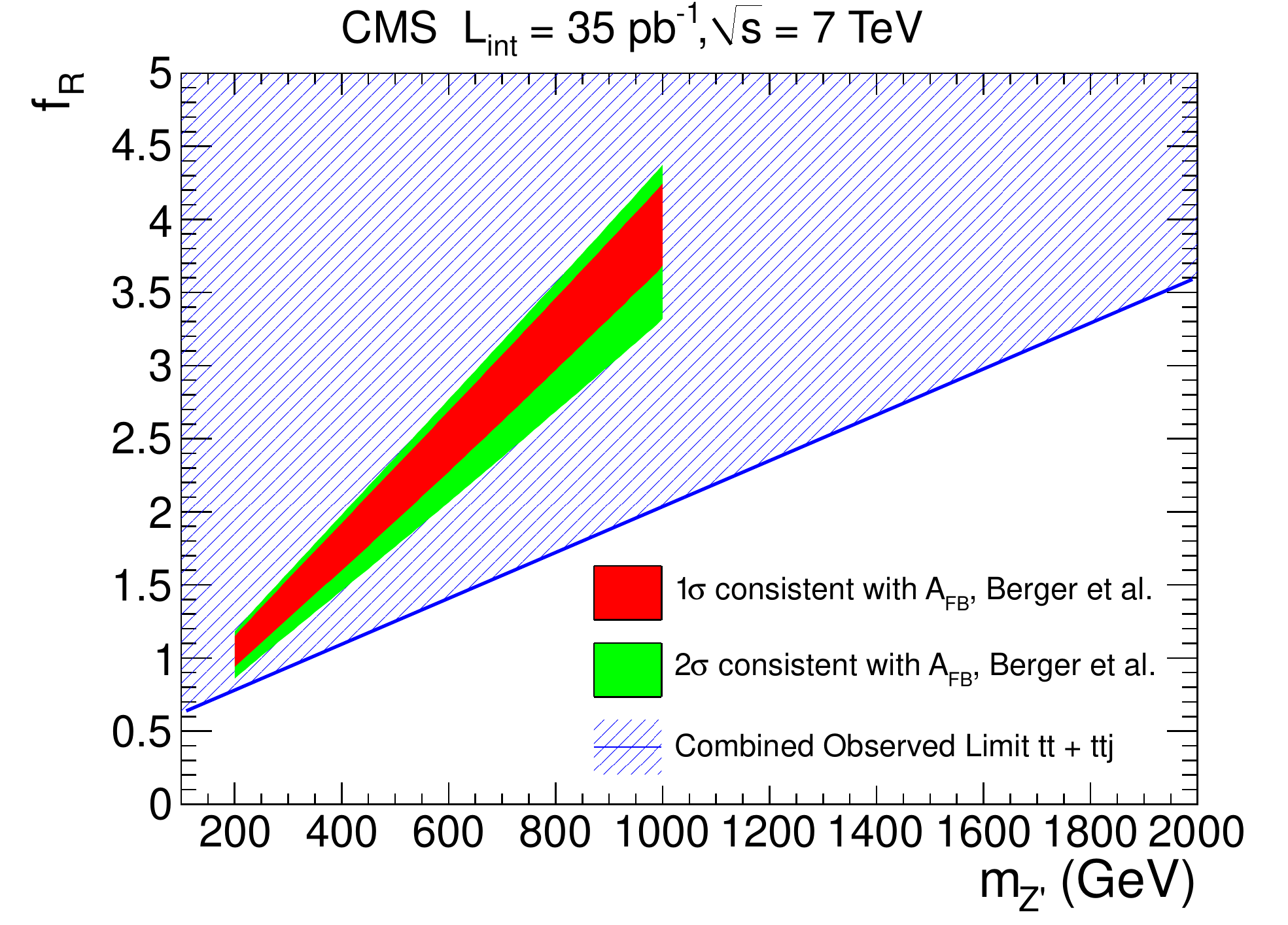} & 
\includegraphics[width=0.45\textwidth]{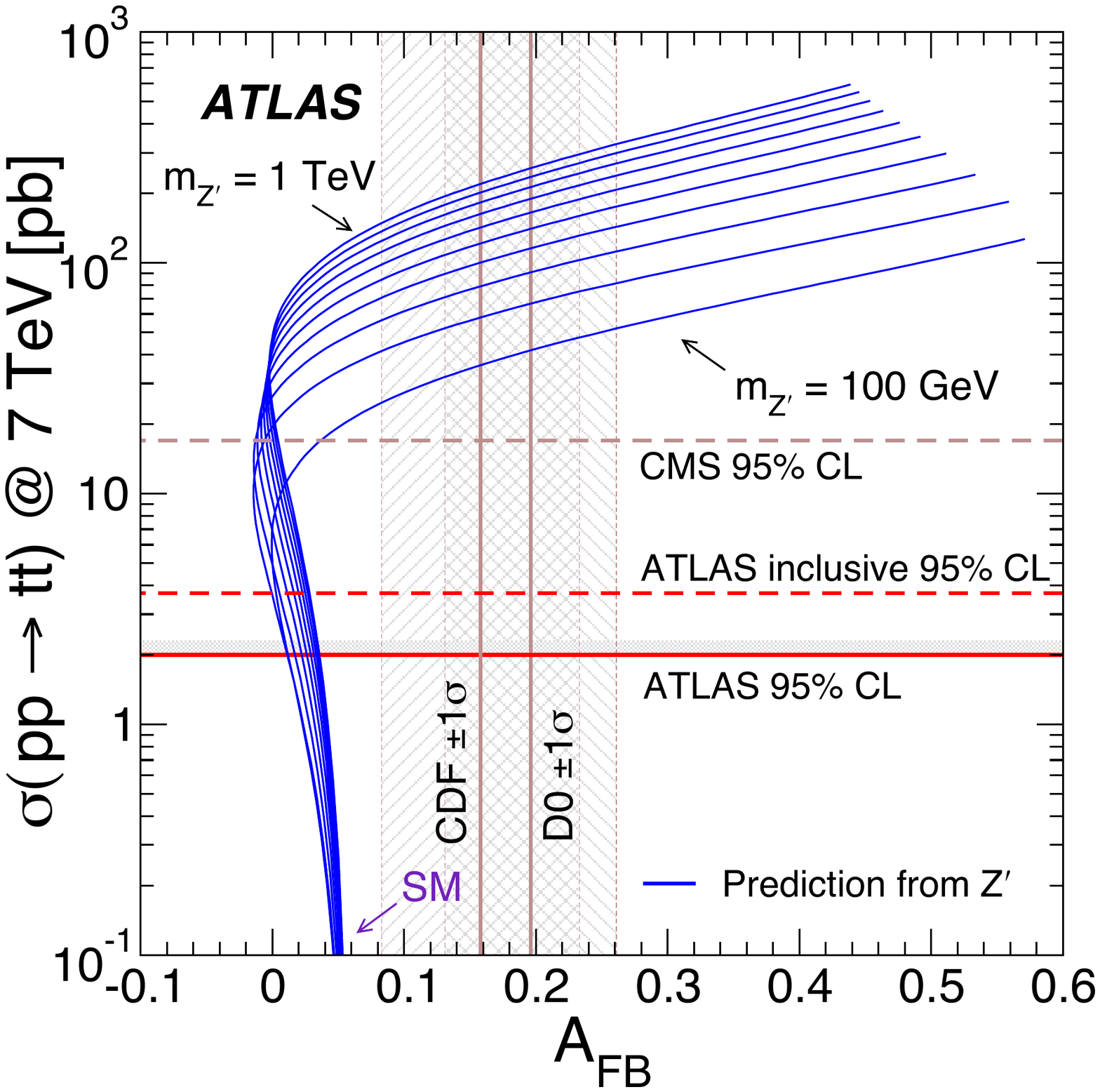} \\
\end{tabular}
\caption{Excluded region shown as the shaded area in the $Z'$ mass vs right handed couplings ($f_R$) frame. One and two standard deviations of region of parameter space consistent with the $A_{FB}$ measurements  at the Tevatron using the leading-order $t\overline{t}$ cross-section calculated by MadGraph are also shown (left plot). Regions consistent with the new physics contributions to the inclusive $A_{FB}$ at the Tevatron, and the $t\overline{t}$ cross-section at LHC (right plot). Limits from the ATLAS analysis at $\sqrt{s}=$7 TeV  are shown by the solid horizontal lines. CDF and D0 $A_{FB}$ measurements are displayed as vertical lines with the associated uncertainty bands (right plot).}\label{FigSameSign3}
\end{center}
\end{figure}
\begin{table}[h]
\centerline{
\begin{tabular}{ccc}
\hline
 Chirality configuration & $\sigma(pp\rightarrow tt)$ [pb] & $|C|/\Lambda^2$ [TeV$^{-2}$] \\
\hline
Left-left & 0.19 & 0.092 \\
Left-right & 0.20 & 0.271 \\
Right-right & 0.21 & 0.099 \\
\hline
 \end{tabular}
}
\caption{Observed limits on the positively-charged $tt$ production.}
\label{tab:samesigntop}
\end{table}


\clearpage
\newpage

\section{Summary and Prospects}
\begin{figure}[h]
\begin{center}
\includegraphics[width=0.5\textwidth]{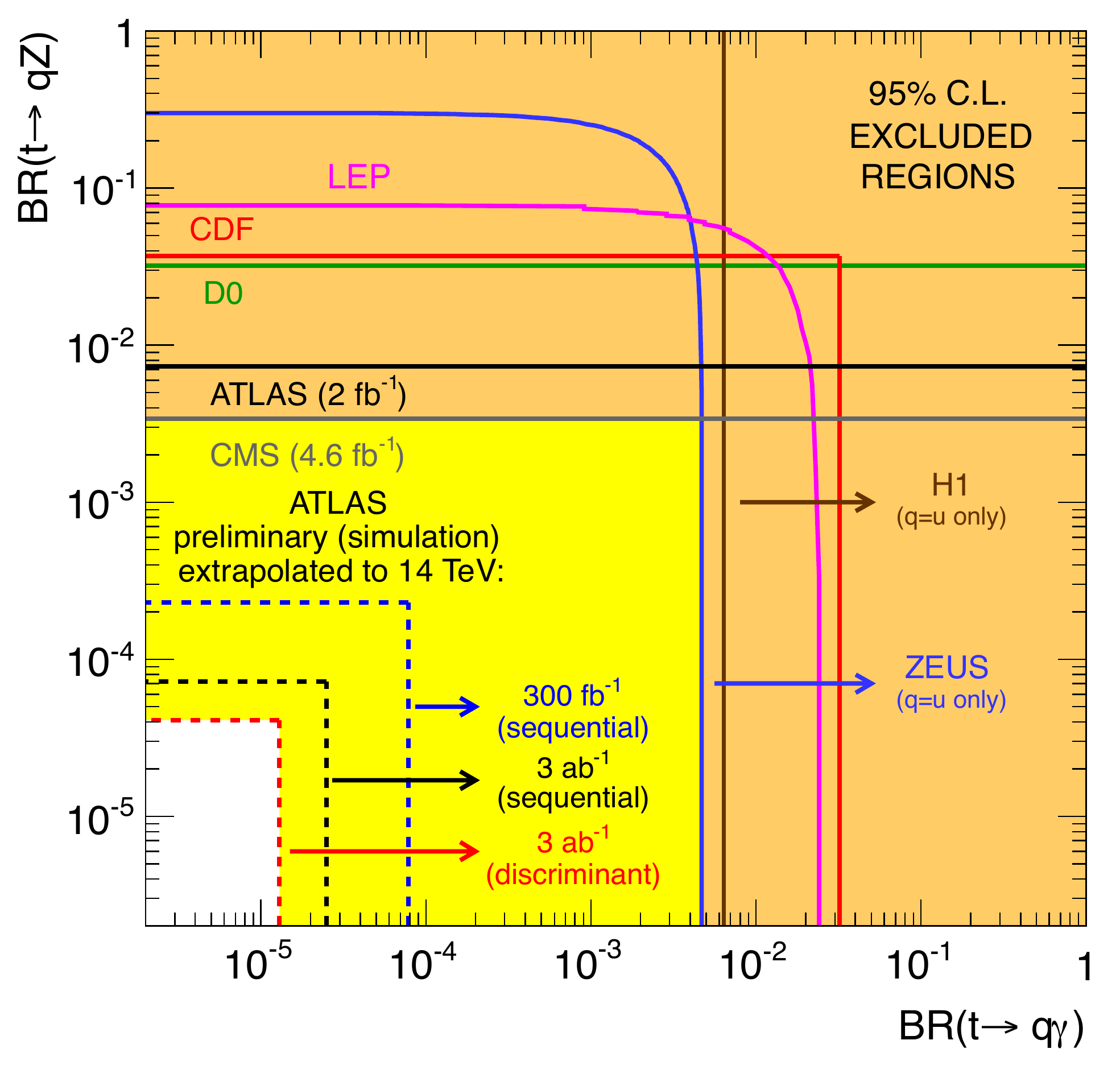}
\caption{The observed 95\% CL limits on the $\mathcal{B}(t\rightarrow q\gamma)$
vs. ($\mathcal{B}(t\rightarrow qZ)$) plane shown with solid lines for the LEP, ZEUS, H1, D0, CDF, ATLAS and CMS collaborations as of August 10, 2012.  The expected sensitivities for ATLAS measurements at 14 TeV with different integrated luminosities are shown with the dashed lines.
}\label{fig:prospects}
\end{center}
\end{figure}

The results of the searches conducted by ATLAS, CDF, CMS, and D0 collaborations are presented. No signs of FCNCs in the decays of $t\overline{t}$, single top quark or same-sign top quark processes have been observed. The exclusion limits are getting closer to the predictions from specific new physics models. First limits on the $t\rightarrow cH$ process are presented in this proceeding are almost at the level of $2HDM$ predictions. At the $\sqrt{s}=$ 14 TeV LHC run, ATLAS and CMS experiments expect the limits to be an order of magnitude smaller as shown in Figure \ref{fig:prospects} in the $\mathcal{B}(t\rightarrow q\gamma)$ vs $\mathcal{B}(t\rightarrow qZ)$ plane. Using 300 fb$^{-1}$ of 14 TeV data, ATLAS collaboration  expects $\mathcal{B}(t\rightarrow qZ)>2\times10^{-4}$ \cite{Aad:2012c} and the CMS experiment expects to exclude $\mathcal{B}(t\rightarrow qZ)>10^{-5}$ \cite{Chatrchyan:2013c}. 

\newpage
\clearpage

%
%
%
%
 

\begin{footnotesize}

\end{footnotesize}


\end{document}